\documentclass[sigconf]{acmart}
\AtBeginDocument{%
  }

\copyrightyear{2025}
\acmYear{2025}
\setcopyright{acmlicensed}\acmConference[CIKM '25]{Proceedings of the 34th ACM International Conference on Information and Knowledge Management}{November 10--14, 2025}{Seoul, Republic of Korea}
\acmBooktitle{Proceedings of the 34th ACM International Conference on Information and Knowledge Management (CIKM '25), November 10--14, 2025, Seoul, Republic of Korea}
\acmDOI{10.1145/3746252.3760909}
\acmISBN{979-8-4007-2040-6/2025/11}

\usepackage{booktabs}
\usepackage{multirow} 
\usepackage{subfigure}
\usepackage{diagbox}
  
\usepackage{amsmath,amsthm,amssymb,amsfonts}
\usepackage{algorithm}
\usepackage{algorithmic}
\usepackage{color, colortbl}

\definecolor{Gray}{rgb}{0.9, 0.9, 0.9}

\begin{document}

\title{Enhancing Graph Collaborative Filtering with FourierKAN Feature Transformation}

\settopmatter{authorsperrow=4}

\author{Jinfeng Xu}
\affiliation{%
  \institution{\small{The University of Hong Kong}}
  \city{Hong Kong}
  \country{China}}
\email{jinfeng@connect.hku.hk}

\author{Zheyu Chen}
\affiliation{%
  \institution{\small{Hong Kong Polytechnic University}}
  \city{Hong Kong}
  \country{China}}
\email{zheyu.chen@connect.polyu.hk}

\author{Jinze Li}
\affiliation{%
  \institution{\small{The University of Hong Kong}}
  \city{Hong Kong}
  \country{China}}
\email{lijinze-hku@connect.hku.hk}

\author{Shuo Yang}
\affiliation{%
  \institution{\small{The University of Hong Kong}}
  \city{Hong Kong}
  \country{China}}
\email{shuoyang.ee@gmail.com}

\author{Wei Wang}
\affiliation{%
    \institution{\small{Shenzhen MSU-BIT University}}  
    \city{Shenzhen}   
    \country{China}}
\email{ehomewang@ieee.org}

\author{Xiping Hu}
\affiliation{%
    \institution{\small{Beijing Institute of Technology}}  
    \city{Beijing}   
    \country{China}}
\email{huxp@bit.edu.cn}

\author{Edith Ngai*}
% \authornote{*Corresponding authors}
\affiliation{%
  \institution{\small{The University of Hong Kong}}
  \city{Hong Kong}
  \country{China}}
\email{chngai@eee.hku.hk}

% 
% \authornote{Corresponding author.}

% \author{Anonymous Authors}
 
\begin{abstract}
Graph Collaborative Filtering (GCF) has emerged as a dominant paradigm in modern recommendation systems, excelling at modeling complex user-item interactions and capturing high-order collaborative signals through graph-structured learning. Most existing GCF models predominantly rely on simplified graph architectures like LightGCN, which strategically remove feature transformation and activation functions from vanilla graph convolution networks. Through systematic analysis, we reveal that feature transformation in message propagation can enhance model representation, though at the cost of increased training difficulty. To this end, we propose FourierKAN-GCF, a novel GCN framework that adopts Fourier Kolmogorov-Arnold Networks as efficient transformation modules within graph propagation layers. This design enhances model representation while decreasing training difficulty. Our FourierKAN-GCF can achieve higher recommendation performance than most widely used GCF backbone models. In addition, it can be integrated into existing advanced self-supervised models as a backbone, replacing their original backbone to achieve enhanced performance. Extensive experiments on three public datasets demonstrate the superiority of FourierKAN-GCF.
\end{abstract}

% <concept_id>10002951.10003317.10003347.10003350</concept_id>
% <concept_desc>Information systems~Recommender systems</concept_desc>

% \ccsdesc[500]{Information systems~Recommender systems;}

\keywords{Recommendation, Collaborative Filtering, Graph Neural Network, Kolmogorov-Arnold Network, Fourier Coefficients}

\maketitle

\section{Introduction}
Recommender systems are widely used to alleviate information overload on the Web \cite{chen2020revisiting,he2017neural,chen2025don,chen2025squeeze,xu2025survey,xu2025mdvt}, aiming to recommend suitable items for users based on their historical behavior. Collaborative filtering (CF) addresses this by learning user preferences from similar users. Recently, graph-based CF (GCF) models \cite{he2020lightgcn,wang2019neural,jiang2023adaptive,xu2025mentor,xu2025best,xu2025cohesion,xu2024improving} have achieved notable success, as user-item interactions are naturally graph-structured. For example, NGCF \cite{wang2019neural} adopts whole standard GCN in recommender systems, which retains the feature transformation and nonlinear operation. However, LightGCN \cite{he2020lightgcn} states that both feature transformation and nonlinear operation are unnecessary in the recommendation field and further proposes a lightweight GCN. Consequently, most subsequent GCF models \cite{chen2020revisiting,cailightgcl,xu2024aligngroup} adopt LightGCN as the backbone for further exploration.

However, there are many unfair analyses of NGCF in LightGCN: a) There are two different feature transformations in NGCF, but LightGCN removed them without performing separate fine-grained ablation experiments. b) LightGCN also removed the interaction information representation part of NGCF without giving reasons. Therefore, we question: \textbf{`Is feature transformation during message passing in GCN really unnecessary in recommendation?'}
% \vskip -0.2in
% \begin{gather*}
%     \textbf{Is feature transformation and nonlinear operation} 
% \\ \textbf{during message passing in GCN}
% \\ \textbf{really unnecessary in recommendation field?}
% \end{gather*}
% \vskip -0.2in

We provide an empirical analysis for NGCF and LightGCN in Section~\ref{pre}. Then, we state that feature transformation in NGCF can enhance the interaction representation and boost the performance of GCN, but increases the training difficulty. In this work, we introduce a simple yet powerful graph-based recommendation model called FourierKAN-GCF. Specifically, FourierKAN-GCF incorporates a unique Fourier Kolmogorov-Arnold Network (KAN) in place of the traditional multilayer perceptron (MLP) within the feature transformation during message passing in GCNs. This substitution enhances the representational capabilities and reduces the difficulty of training for GCFs. FourierKAN-GCF can achieve higher recommendation performance than most widely-used GCF backbone models. In addition, FourierKAN-GCF can be integrated into existing advanced self-supervised models as a backbone, replacing their original backbone to achieve enhanced performance. Extensive experiments on public datasets demonstrate the superiority of FourierKAN-GCF over state-of-the-art methods. Our work is intended to reawaken researchers' thinking about feature transformation in GCF, rather than arbitrarily removing it.

\section{Preliminary}
\label{pre}
% We first introduce two popular graph-based collaborative filtering methods: NGCF and LightGCN. 

\subsection{NGCF Brief}
NGCF \cite{wang2019neural} retains feature transformation and nonlinear operation during the message passing in GCN, formally:
% \begin{equation}
% \label{eq:ngcf1}
% \begin{aligned} \mathbf{e}_u^{(l+1)} & =\sigma(\overbrace{\mathbf{W}_1 \mathbf{e}_u^{(l)}}^{\text{Ego Rep.}}+\sum_{i \in \mathcal{N}_u} \frac{\overbrace{\mathbf{W}_1 \mathbf{e}_i^{(l)}}^{\text{Neighbors Rep.}}+\overbrace{\mathbf{W}_2(\mathbf{e}_i^{(l)} \odot \mathbf{e}_u^{(l)})}^{\text{Interactions Rep.}}}{\sqrt{|\mathcal{N}_u||\mathcal{N}_i|}}), \\ \mathbf{e}_i^{(l+1)} & =\sigma(\overbrace{\mathbf{W}_1 \mathbf{e}_i^{(l)}}^{\text{Ego Rep.}}+\sum_{u \in \mathcal{N}_i} \frac{\overbrace{\mathbf{W}_1 \mathbf{e}_u^{(l)}}^{\text{Neighbors Rep.}}+\overbrace{\mathbf{W}_2(\mathbf{e}_u^{(l)} \odot \mathbf{e}_i^{(l)})}^{\text{Interactions Rep.}}}{\sqrt{|\mathcal{N}_u||\mathcal{N}_i|}}),
% \end{aligned}
% \end{equation}
\vskip -0.2in
\begin{equation}
\label{eq:ngcf1}
\begin{aligned} \mathbf{e}_u^{(l+1)} & =\sigma(\mathbf{W}_1 \mathbf{e}_u^{(l)}+\sum_{i \in \mathcal{N}_u} \frac{\mathbf{W}_1 \mathbf{e}_i^{(l)}+\mathbf{W}_2(\mathbf{e}_i^{(l)} \odot \mathbf{e}_u^{(l)})}{\sqrt{|\mathcal{N}_u||\mathcal{N}_i|}}), \\ \mathbf{e}_i^{(l+1)} & =\sigma(\mathbf{W}_1 \mathbf{e}_i^{(l)}+\sum_{u \in \mathcal{N}_i} \frac{\mathbf{W}_1 \mathbf{e}_u^{(l)}+\mathbf{W}_2(\mathbf{e}_u^{(l)} \odot \mathbf{e}_i^{(l)})}{\sqrt{|\mathcal{N}_u||\mathcal{N}_i|}}),
\end{aligned}
\end{equation}
\vskip -0.12in
\noindent where $\mathbf{e}_u^{(l)}$ and $\mathbf{e}_i^{(l)}$ represent the embedding of user $u$ and item $i$ after $l$ layers message propagation, respectively. $\mathcal{N}_u$ and $\mathcal{N}_i$ denote the interacted item set with $u$ and interacted user set with $i$, respectively. $\mathbf{W}_1$ and $\mathbf{W}_2$ are two trainable weight matrices to perform feature transformation in each layer. $\sigma$ is the nonlinear activation function. It is worth noting that $\mathbf{W}_1 \mathbf{e}_i^{(l)}$ and $\mathbf{W}_1 \mathbf{e}_u^{(l)}$ can be regarded as the aggregated representation from neighbors, while $\mathbf{W}_2(\mathbf{e}_i^{(l)} \odot \mathbf{e}_u^{(l)})$ and $\mathbf{W}_2(\mathbf{e}_u^{(l)} \odot \mathbf{e}_i^{(l)})$ can be regarded as the aggregated representation from interaction information. 
The final embeddings are calculated by $\mathbf{\hat{e}}_u = \{\mathbf{e}_u^{(1)}||\mathbf{e}_u^{(2)}||...||\mathbf{e}_u^{(L)}\}$ and $
\mathbf{\hat{e}}_i = \{\mathbf{e}_i^{(1)}||\mathbf{e}_i^{(2)}||...||\mathbf{e}_0^{(L)}\}$, where $||$ is the concatenation operation. NGCF only concatenates layer-$1$ to layer-$L$ and ignores ego layer-$0$, since ego layer-$0$ has already been considered in the first term $\mathbf{W}_1 \mathbf{e}_u^{(l)}$ and $\mathbf{W}_1 \mathbf{e}_i^{(l)}$ of message passing and propagation.
% \vskip -0.2in
% \begin{equation}
% % \begin{aligned}
% \mathbf{\hat{e}}_u = \mathbf{e}_u^{(1)}||\mathbf{e}_u^{(2)}||...||\mathbf{e}_u^{(L)}, \quad
% \mathbf{\hat{e}}_i = \mathbf{e}_i^{(1)}||\mathbf{e}_i^{(2)}||...||\mathbf{e}_0^{(L)},
% % \end{aligned}
% \end{equation}
% \vskip -0.1in
\subsection{LightGCN Brief}
LightGCN \cite{he2020lightgcn} analyzes the feature transformation and nonlinear operation in NGCF. It offers four observations: a) Removing the entire feature transformation, including $\mathbf{W}_1$ and $\mathbf{W}_2$, leads to consistent improvements over NGCF. b) Removing only nonlinear operation $\sigma$ will lead to a small deterioration of performance. c) Removing both entire feature transformation and nonlinear operation can improve performance significantly. d) The deterioration of NGCF stems from the training difficulty rather than over-fitting.

To this end, LightGCN removes feature transformation and nonlinear operation during the message passing in GCN. Formally, the user-item graph to propagate embeddings as:
% \begin{equation}
% \label{eq:lightgcn1}
% \mathbf{e}_u^{(l+1)} =\sum_{i \in \mathcal{N}_u} \frac{\overbrace{\mathbf{e}_i^{(l)}}^{\text{Neighbors Rep.}}}{\sqrt{|\mathcal{N}_u||\mathcal{N}_i|}}, \quad \mathbf{e}_i^{(l+1)} =\sum_{u \in \mathcal{N}_i} \frac{\overbrace{\mathbf{e}_u^{(l)}}^{\text{Neighbors Rep.}}}{\sqrt{|\mathcal{N}_u||\mathcal{N}_i|}},
% \end{equation}
\vskip -0.15in
\begin{equation}
\label{eq:lightgcn1}
\mathbf{e}_u^{(l+1)} =\sum_{i \in \mathcal{N}_u} \frac{\mathbf{e}_i^{(l)}}{\sqrt{|\mathcal{N}_u||\mathcal{N}_i|}}, \quad \mathbf{e}_i^{(l+1)} =\sum_{u \in \mathcal{N}_i} \frac{\mathbf{e}_u^{(l)}}{\sqrt{|\mathcal{N}_u||\mathcal{N}_i|}},
\end{equation}
\vskip -0.1in
\noindent where $\mathbf{e}_u^{(l)}$ and $\mathbf{e}_i^{(l)}$ represent the embedding of user $u$ and item $i$ after $l$ layers message propagation, respectively. $\mathcal{N}_u$ and $\mathcal{N}_i$ denote the interacted item set with $u$ and interacted user set with $i$, respectively. LightGCN only retains the aggregation of neighbors' representation but removes the aggregation of interaction information representation. The final embeddings are calculated as $\mathbf{\hat{e}}_u = \sum_{l=0}^{L}\frac{1}{L + 1}\mathbf{e}_u^{(l)}$ and $\mathbf{\hat{e}}_i = \sum_{l=0}^{L}\frac{1}{L + 1}\mathbf{e}_i^{(l)}$, where $L$ is the total layer number. LightGCN considers the ego layer-$0$ cause it removes the first term $\mathbf{W}_1 \mathbf{e}_u^{(l)}$ and $\mathbf{W}_1 \mathbf{e}_i^{(l)}$. 
% \vskip -0.1in
% \begin{equation}
% % \begin{aligned}
% \mathbf{\hat{e}}_u = \sum_{l=0}^{L}\frac{\mathbf{e}_u^{(l)}}{L + 1}, \quad
% \mathbf{\hat{e}}_i = \sum_{l=0}^{L}\frac{\mathbf{e}_i^{(l)}}{L + 1},
% % \end{aligned}
% \end{equation}
% \vskip -0.05in

We argue that observation (d) is the main reason why LightGCN performs better than NGCF in most cases. Observations (a-c) were not sufficient to verify that the entire feature transformation and nonlinear operation do not contribute to feature extraction. Besides, we argue that LightGCN verified the feature transformation not contributing to learning better features by comparing NGCF with removed both $\mathbf{W}_1$ and the whole interaction representation aggregation part $\mathbf{W}_2(\mathbf{e}_i^{(l)} \odot \mathbf{e}_u^{(l)})$, which is not a fair comparison. We point out that $\mathbf{e}_u^{(l)}$ and $\mathbf{e}_i^{(l)}$ naturally contain information about user preferences and item properties that can be adequately described through multiple feature dimensions. Therefore, $\mathbf{W}_1$ is an unnecessary feature transformation part. However, we point out that the interaction representation aggregation part $\mathbf{W}_2(\mathbf{e}_i^{(l)} \odot \mathbf{e}_u^{(l)})$ contains valuable interaction information, which can not be easily extracted by the heuristic rule. In this case, $\mathbf{W}_2$ is a necessary feature transformation part that contributes to feature extraction. In the next subsection, we provide an empirical analysis.

% \vskip -0.1in
% \begin{table}[!ht]
%     \centering
% \caption{Performance comparison of LightGCN, NGCF, and four variants of NGCF in terms of Recall@20 and NDCG@20.}
% \vskip -0.15in
% \label{tab:1}
% \resizebox{\linewidth}{!}{
%     \begin{tabular}{c|c|c|ccccc}
%     \hline
%          Dataset&  Metrics&  LightGCN&  NGCF&  NGCF-f1&  NGCF-f2 & NGCF-i & NGCF-n\\
%          \hline
%          \multirow{2}{*}{MOOC} & Recall@20 & 0.3307 & 0.3361 & 0.3377$\uparrow$ & 0.3357$\downarrow$ & 0.3301$\downarrow$ & 0.3343$\downarrow$ \\
%          & NDCG@20 & 0.1811 & 0.1894 & 0.1926$\uparrow$ & 0.1897$\uparrow$ & 0.1824$\downarrow$ & 0.1878$\downarrow$ \\
%          \hline
%          \multirow{2}{*}{Amazon} & Recall@20 & 0.0447 & 0.0379 & 0.0414$\uparrow$ & 0.0388$\uparrow$ & 0.0362$\downarrow$ & 0.0373$\downarrow$ \\
%          & NDCG@20 & 0.0227 & 0.0196 & 0.0209$\uparrow$ & 0.0192$\downarrow$ & 0.0181$\downarrow$ & 0.0196$-$ \\
%         \hline
%          \multirow{2}{*}{Gowalla} & Recall@20 & 0.1830 & 0.1755 & 0.1791$\uparrow$ & 0.1764$\uparrow$ & 0.1739$\downarrow$ & 0.1750$\downarrow$ \\
%          & NDCG@20 & 0.1152 & 0.1013 & 0.1081$\uparrow$ & 0.1020$\uparrow$ & 0.1010$\downarrow$ & 0.1008$\downarrow$ \\
%          \hline
%     \end{tabular}
%     }
%      \vskip -0.2in
% \end{table}
% \vskip -0.2in

\vskip -0.1in
\begin{table}[!ht]
    \centering
\caption{Performance comparison of LightGCN, NGCF, and six variants of NGCF in terms of Recall@20 (R@20) and NDCG@20 (N@20).}
\vskip -0.15in
\label{tab:1}
\resizebox{\linewidth}{!}{
    \begin{tabular}{c|cccccc}
         \toprule
         \multirow{1}{*}{\textbf{Datasets}}&  \multicolumn{2}{c}{\textbf{MOOC}}&\multicolumn{2}{c}{\textbf{Amazon}}& \multicolumn{2}{c}{\textbf{Gowalla}}\\
         \midrule
         \multirow{1}{*}{\textbf{Metrics}} & R@20 & N@20 & R@20 & N@20& R@20 & N@20 \\ \midrule
         LightGCN& 0.3307& 0.1811& 0.0447& 0.0227& 0.1830& 0.1152\\\midrule
         NGCF& 0.3361& 0.1894& 0.0379& 0.0196& 0.1755& 0.1013\\
         NGCF-f1& 0.3377$\uparrow$ & 0.1926$\uparrow$ & 0.0414$\uparrow$ & 0.0209$\uparrow$ &0.1791$\uparrow$ &0.1081$\uparrow$\\
         NGCF-f2& 0.3357$\downarrow$ & 0.1897$\uparrow$& 0.0388$\uparrow$ & 0.0192$\downarrow$ & 0.1764$\uparrow$ & 0.1020$\uparrow$ \\
         NGCF-i& 0.3301$\downarrow$ & 0.1824$\downarrow$ & 0.0362$\downarrow$ & 0.0181$\downarrow$ & 0.1739$\downarrow$ & 0.1010$\downarrow$ \\
         NGCF-f1-f2& 0.3374$\uparrow$ & 0.1913$\uparrow$ & 0.0407$\uparrow$ & 0.0205$\uparrow$ &0.1784$\uparrow$ &0.1077$\uparrow$\\
         NGCF-f1-i& 0.3332$\downarrow$ & 0.1868$\downarrow$ & 0.0372$\downarrow$ & 0.0190$\downarrow$ &0.1752$\downarrow$ &0.1013$-$\\
         NGCF-n& 0.3343$\downarrow$ & 0.1878$\downarrow$ & 0.0373$\downarrow$ & 0.0196$-$ & 0.1750$\downarrow$ & 0.1008$\downarrow$ \\  
         \bottomrule
    \end{tabular}
    }
    \vskip -0.1in
\end{table}

\subsection{Re-Analysis for NGCF}
We implement six variants of NGCF: 1) NGCF-f1 removes feature transformation matrix $\mathbf{W}_1$. 2) NGCF-f2 removes feature transformation matrix $\mathbf{W}_2$. 3) NGCF-i removes the whole interaction representation aggregation part $\mathbf{W}_2(\mathbf{e}_i^{(l)} \odot \mathbf{e}_u^{(l)})$. 4) NGCF-f1-f2 removes both $\mathbf{W}_1$ and $\mathbf{W}_2$. 5) NGCF-f1-i removes both $\mathbf{W}_1$ and $\mathbf{W}_2(\mathbf{e}_i^{(l)} \odot \mathbf{e}_u^{(l)})$. 6) NGCF-n removes the nonlinear operation $\sigma$.

We keep all optimal hyper-parameter settings as the NGCF reported. As Table~\ref{tab:1} shows, we conclude the findings: a) For all three datasets, removing the feature transformation matrix $\mathbf{W}_1$ will lead to observed improvements. Therefore, the feature transformation matrix $\mathbf{W}_1$ is unnecessary. b) For the MOOC dataset, removing the feature transformation matrix $\mathbf{W}_2$ leads to a slight deterioration. However, for the other two datasets, it will lead to a small improvement. Moreover, removing both $\mathbf{W}_1$ and $\mathbf{W}_2$ will lead to a small deterioration than only removing $\mathbf{W}_1$ for all datasets. We own this phenomenon to that removing only $\mathbf{W}_1$ can make the model optimization focus on training $\mathbf{W}_2$, and remove both $\mathbf{W}_1$ and $\mathbf{W}_2$ will lose the representation power from feature transformation. c) Removing the whole interaction representation aggregation part $\mathbf{W}_2(\mathbf{e}_i^{(l)} \odot \mathbf{e}_u^{(l)})$ will lead to a performance degradation on all datasets. Additionally, removing both $\mathbf{W}_1$ and $\mathbf{W}_2(\mathbf{e}_i^{(l)} \odot \mathbf{e}_u^{(l)})$ will make a obviously degration than only remove $\mathbf{W}_1$ and even NGCF. This shows that this part can bring valuable information about interaction. d) Nonlinear operation only brings a small positive effect. e) NGCF outperforms LightGCN on the MOOC dataset verifying that NGCF has better representation power than LightGCN, but the training is more difficult \cite{he2020lightgcn}. Then, we can draw some conclusions. First, the feature transformation matrix $\mathbf{W}_1$ is unnecessary for NGCF. Besides, the feature transformation matrix $\mathbf{W}_2$ and the whole interaction representation aggregation part $\mathbf{W}_2(\mathbf{e}_i^{(l)} \odot \mathbf{e}_u^{(l)})$ is beneficial for feature extraction. Last, nonlinear operation only has a minor influence. 

To this end, a feature transformation approach, simpler to train than MLPs but with strong representational capabilities, could be effective in graph learning for recommendations. KAN \cite{liu2024kan} is regarded as a promising alternative to MLP. While KAN shares the same theoretically unlimited representational capacity as MLP \cite{wang2024expressiveness}, the practical representational capacity of MLP is constrained by the hidden dimensionality. In contrast, the practical representational capacity of KAN depends on its ability to fit trainable activation functions. Our work can be seen as exploring adopting KAN as a feature transformation approach within GCF.

% To this end, we believe that a feature transformation approach, simpler to train than MLPs but with strong representational capabilities, could be effective in graph learning for recommendations. The novel Kolmogorov-Arnold Network model \cite{liu2024kan}, with its MLP-comparable feature extraction abilities and easier training difficulty, fits this need well.

\section{Methodology}
\begin{figure}
    \centering
    \includegraphics[width=1\linewidth]{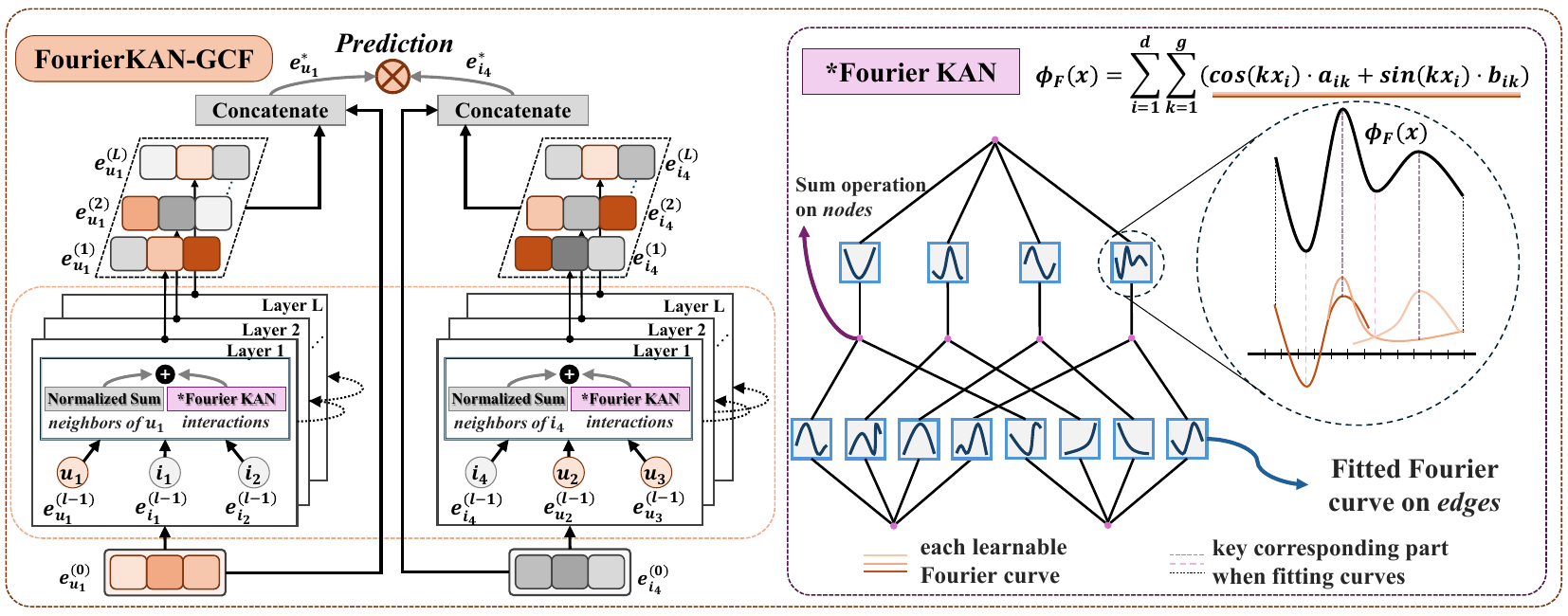}
    \vskip -0.15in
    \caption{Overview of FourierKAN-GCF.}
    \label{FourierKAN-GCF}
    \vskip -0.2in
\end{figure}

We detail the overview of FourierKAN-GCF\footnote[1]{Code is available at: \href{https://anonymous.4open.science/r/FourierKAN-GCF-r}{https://anonymous.4open.science/r/FourierKAN-GCF-r}.} in Figure~\ref{FourierKAN-GCF}. FourierKAN-GCF introduces a promising feature transformation for GCF, which significantly boosts performance and simplifies training. Note that FourierKAN-GCF can be adopted as the backbone in existing advanced self-supervised models. 

% \begin{figure*}
%     \centering
%     \includegraphics[width=0.9\linewidth]{pic/FourierKAN-GCF.pdf}
%     \vskip -0.15in
%     \caption{Overview of FourierKAN-GCF.}
%     \label{FourierKAN-GCF}
%     \vskip -0.2in
% \end{figure*}

\subsection{Kolmogorov-Arnold Network (KAN)}
Kolmogorov-Arnold Network \cite{liu2024kan} is a promising alternative to Multi-Layer Perceptron (MLP). MLP is inspired by the universal approximation theorem \cite{cybenko1989approximation}. KAN focuses on the Kolmogorov-Arnold representation theorem 
\cite{kolmogorov1961representation}. Specifically, unlike MLPs, which have fixed activation functions on nodes, KANs contain learnable activation functions on edges (weights). This unique architecture enables KANs to learn nonlinear functions more effectively. Formally:
\vskip -0.15in
\begin{equation}
    \operatorname{KAN} = f(\mathbf{x})=\sum_{q=1}^{2 n+1} \Phi_q(\sum_{p=1}^n \phi_{q, p}(\mathbf{x}_p)),
\end{equation}
\vskip -0.05in
\noindent where $\phi_{q, p}$ are univariate functions that map each input variable $\mathbf{x}_p$ such
$\phi_{q, p}:[0,1] \rightarrow \mathbb{R}$ and $\phi_{q}: \mathbb{R} \rightarrow \mathbb{R}$. $\phi_{q, p}(\mathbf{x}_p)$ is trainable activation function. In the KolmogovArnold theorem, the inner functions form a KAN layer with $n_{in}$ = $n$ and
$n_{out}$ = $2n + 1$, and the outer functions form a KAN layer with $n_{in}$ = $2n + 1$
and $n_{out}$ = $n$. So the Kolmogorov-Arnold representations in this formula are simply
compositions of two KAN layers. A useful trick is that it includes a basis function $b(\mathbf{x})$ such that the activation function $\operatorname{silu}(\mathbf{x})=\frac{x}{1+e^{-\mathbf{x}}}$. $\phi(\mathbf{x})=\mathbf{w}(b(\mathbf{x})+\operatorname{spline}(\mathbf{x}))$ is the sum of the basis function $b(\mathbf{x})$ and function $\operatorname{spline}(\mathbf{x})=\sum_i c_i B_i(\mathbf{x})$ is a linear combination of B-splines, where $c_i$ is a trainable parameter.
% \vskip -0.15in
% \begin{equation}
% \phi(\mathbf{x})=\mathbf{w}(b(\mathbf{x})+\operatorname{spline}(\mathbf{x})),
% \end{equation}
% \vskip -0.25in
% % \vskip -0.1in
% \begin{equation}
% b(x)=\operatorname{silu}(\mathbf{x})=\frac{x}{1+e^{-\mathbf{x}}}, \quad
% \operatorname{spline}(\mathbf{x})=\sum_i c_i B_i(\mathbf{x}),
% \end{equation}
% \vskip -0.1in
% \noindent 

\subsection{Fourier KAN}
To further reduce the training difficulty and be able to adopt to different scenarios. Our goal can be converted into finding the split from a complex function into multiple relatively simple nonlinear functions. Naturally, the Fourier Coefficients \cite{nussbaumer1982fast} is a potential choice. Therefore, we propose the following equation:
\vskip -0.15in
\begin{equation}
\phi_F(\mathbf{x})=\sum_{i=1}^{d} \sum_{k=1}^{g}\left(\cos \left(k \mathbf{x}_i\right) \cdot a_{i k}+\sin \left(k \mathbf{x}_i\right) \cdot b_{i k}\right),
\end{equation}
\vskip -0.05in
\noindent where $d$ is the dimension number of features. Fourier coefficients $a_{i k}$ and $b_{i k}$ are trainable. Hyper-parameter $g$ is the gridsize, which plays a critical role in controlling the number of terms (frequencies) used in the Fourier series expansion. Specifically, $g$ determines how many different sine and cosine terms are included in the Fourier Coefficients corresponding to each input dimension. The Fourier Coefficients has a significant advantage in computational efficiency and reduces the training difficulty caused by the spline function.

\subsection{FourierKAN-GCF}
The message passing in FourierKAN-GCF is defined as:
\vskip -0.15in
\begin{equation}
\label{eq:fkan-gcf}
\begin{aligned} 
\mathbf{e}_u^{(l+1)} & =\sigma(\mathbf{e}_u^{(l)}+\sum_{i \in \mathcal{N}_u} \frac{\mathbf{e}_i^{(l)}+\phi_F(\mathbf{e}_i^{(l)} \odot \mathbf{e}_u^{(l)})}{\sqrt{|\mathcal{N}_u||\mathcal{N}_i|}}), \\ \mathbf{e}_i^{(l+1)} & =\sigma(\mathbf{e}_i^{(l)}+\sum_{u \in \mathcal{N}_i} \frac{\mathbf{e}_u^{(l)}+\phi_F(\mathbf{e}_u^{(l)} \odot \mathbf{e}_i^{(l)})}{\sqrt{|\mathcal{N}_u||\mathcal{N}_i|}}),
\end{aligned}
\end{equation}
\vskip -0.05in
\noindent where $\phi_F(\cdot)$ is simplified single layer Fourier KAN function. We remove the unnecessary transform matrix $\mathbf{W}_1$ in NGCF and utilize our Fourier KAN function to replace the transform matrix $\mathbf{W}_2$. The final user embedding and item embedding are calculated by $\mathbf{\hat{e}}_u = \{\mathbf{e}_u^{(1)}||\mathbf{e}_u^{(2)}||...||\mathbf{e}_u^{(L)}\}$ and $
\mathbf{\hat{e}}_i = \{\mathbf{e}_i^{(1)}||\mathbf{e}_i^{(2)}||...||\mathbf{e}_0^{(L)}\}$, where $||$ is the concatenation operation.
% \vskip -0.2in
% \begin{equation}
% % \begin{aligned}
% \mathbf{\hat{e}}_u = \mathbf{e}_u^{(1)}||\mathbf{e}_u^{(2)}||...||\mathbf{e}_u^{(L)}, \quad
% \mathbf{\hat{e}}_i = \mathbf{e}_i^{(1)}||\mathbf{e}_i^{(2)}||...||\mathbf{e}_i^{(L)},
% % \end{aligned}
% \end{equation}
% \vskip -0.05in

\subsection{Dropout Strategies}
To mitigate overfitting in FourierKAN-GCF, we employ message dropout and node dropout strategies, similar to NGCF \cite{wang2019neural}. In message dropout, a fraction $1-p_m$ of message passing in Eq.~\ref{eq:fkan-gcf} is randomly set to zero, where $p_m$ is the dropout ratio. In node dropout, $1-p_n$ of the nodes in the matrix are randomly dropped, with $p_n$ as the dropout ratio.

\subsection{Model Training}
For model optimization, we adopt the Bayesian Personalized Ranking (BPR) \cite{rendle2009bpr} loss function as our optimization criterion. The core objective of BPR is to enhance the divergence in the predictive preference between positive and negative items within each user-item triplet $(u, i_p, i_n) \in \mathcal{D}$, where $\mathcal{D}$ signifies the collection of training data, the term positive item $p$ pertains to an item with which the user $u$ has interacted, whereas the negative item $n$ is selected randomly from the pool of items without interaction with user $u$.
% \vskip -0.1in
\vskip -0.12in
\begin{equation}
\mathcal{L}=\sum_{(u, i_p, i_n) \in \mathcal{D}} - \ln \sigma(\mathbf{\hat{e}}_u^{T}\mathbf{\hat{e}}_{i_p}-\mathbf{\hat{e}}_u^{T}\mathbf{\hat{e}}_{i_n})+\lambda\|\mathbf{\Theta}\|^2,
\end{equation}
\vskip -0.05in
% \vskip -0.1in
\noindent where $\lambda$ controls the $L_2$ regularization strength, $\sigma$ is the Sigmoid function, and $\mathbf{\Theta}$ denotes model parameters.

\section{Experiment}
In this section, we compare the performance of FourierKAN-GCF with popular graph-based backbone models and demonstrate its compatibility with advanced self-supervised graph-based models.

\begin{table}
    \centering
    % \vskip -0.1in
\caption{Statistics of experimental datasets.}
\vskip -0.15in
\label{tab:dataset_statistics}
    \begin{tabular}{ccccc}
    \hline
         Dataset&  \# Users&  \# Items&  \# Interaction& Sparsity\\
         \hline
         MOOC &  82,535 &  1,302 &  458,453 & 99.57\%\\
         Amazon &  50,677 &  16,897 &  454,529 & 99.95\%\\
         Gowalla & 29,859 & 40,989 & 1,027,464 & 99.92\%\\
         \hline
    \end{tabular}
    \vskip -0.2in
\end{table}

\subsection{Datasets and Evaluation Metrics}
\begin{table*}[!t]
\vskip -0.1in
    \centering
\caption{Performance comparison of baselines and FourierKAN-GCF in terms of R@K and N@K. The superscript $^*$ indicates the improvement is statistically significant where the p-value is less than 0.05. \#T denotes seconds per epoch.}
\vskip -0.15in
\label{tab:results}
\resizebox{1\linewidth}{!}{
    \begin{tabular}{c|ccccc|ccccc|ccccc}
    \hline
          Datasets&  \multicolumn{5}{c}{MOOC}& \multicolumn{5}{c}{Amazon} & \multicolumn{5}{c}{Gowalla}\\\hline
          Metrics& R@20& R@50& N@20& N@50& \#T& R@20& R@50& N@20& N@50& \#T& R@20& R@50& N@20& N@50& \#T \\\hline
          % BPR-MF & 0.3353 & 0.4813 & 0.1898 & 0.2261 & & 0.0369 & 0.0699 & 0.0183 & 0.0265 & & 0.1695 & 0.2756 & 0.0988 & 0.1250 &\\
          % BUIR & 0.3196 & 0.4801 & 0.1835 & 0.2224 & & 0.0384 & 0.0749 & 0.0192 & 0.0282 & & 0.1737& 0.2813& 0.1002& 0.1261 &\\
          NGCF & 0.3361 & 0.4799 & 0.1894 & 0.2349 & 5.2s & 0.0379 & 0.0782 & 0.0196 & 0.0274 & 4.4s & 0.1755& 0.2811& 0.1013& 0.1270 & 37.5s\\
          LR-GCCF & 0.3336 & 0.4809 & 0.1938 & 0.2294 & 5.1s & 0.0440 & 0.0815 & 0.0224 & 0.0317 & 4.7s & 0.1803& 0.2971& 0.1101& 0.1369 & 40.1s\\
          LightGCN & 0.3307 & 0.4773 & 0.1811 & 0.2217 & \textbf{4.3s} & 0.0447 & 0.0844 & 0.0227 & 0.0326 & \textbf{3.5s} & 0.1830 & \underline{0.3044} & \underline{0.1152} & \underline{0.1414} & \textbf{26.4s} \\
          UltraGCN & 0.3194 & 0.4701 & 0.1962 & 0.2307 & 4.9s & 0.0459 & \underline{0.0844} & 0.0230 & \underline{0.0331} & 4.0s & 0.1798& 0.2909& 0.1059& 0.1328 & 33.8s\\
          IMP-GCN & 0.2788 & 0.4183 & 0.1717 & 0.2057 & 54.2s & \underline{0.0461} & 0.0839 & \underline{0.0232} & 0.0323 & 44.9s & 0.1808& 0.2932& 0.1060& 0.1345 & 143.6s\\ 
          % LightGCL & 0.2742 & 0.4139 & 0.1697 & 0.2018 & 37.0s & 0.0453 & 0.0818 & 0.0225 & 0.0318 & 29.6s & 0.1888& 0.2976& 0.1081& 0.1348 &87.9s \\ \hline
          KAN-GCF & \underline{0.3417} & \underline{0.4984} & \underline{0.2024} & \underline{0.2396} & 4.9s & 0.0451 & 0.0837 & 0.0229 & 0.0325 & 4.0s & \underline{0.1922} & 0.3023 & 0.1142& 0.1403 & 32.9s\\ 
          \textbf{FourierKAN-GCF} & \textbf{0.3564} & \textbf{0.5065} & \textbf{0.2147} & \textbf{0.2462} & \underline{4.6s} & \textbf{0.0473} & \textbf{0.0856} & \textbf{0.0252} & \textbf{0.0342} & \underline{3.7s} & \textbf{0.1962}& \textbf{0.3077}& \textbf{0.1179}& \textbf{0.1436} & \underline{29.8s}
          \\\hline
          w/o MD & 0.3523 & 0.4912 & 0.2116 & 0.2449 & 4.5s & 0.0452 & 0.0825 & 0.0221 & 0.0314 & 3.7s & 0.1920& 0.3031& 0.1150& 0.1408 & 29.5s\\
          w/o ND & 0.3527 & 0.4839 & 0.2071 & 0.2439 & 4.5s & 0.0452 & 0.0809 & 0.0219 & 0.0309 & 3.6s & 0.1908& 0.3017& 0.1133& 0.1389 & 29.6s\\
          % \hline
          \hline
    \end{tabular}
    }
    % \\ $^1$Datasets can be accessed at \href{http://moocdata.cn}{http://moocdata.cn}, \href{http://jmcauley.ucsd.edu/data/amazon/links.html}{http://jmcauley.ucsd.edu/data/amazon/links.html}, and \href{https://snap.stanford.edu/data/loc-gowalla.html}{https://snap.stanford.edu/data/loc-gowalla.html}.
\vskip -0.1in
\end{table*}

\begin{table*}[!t]
\vskip -0.05in
    \centering
\caption{Performance comparison of different backbones on advanced recommendation models in terms of R@K and N@K.}
\vskip -0.15in
\label{tab:results2}
\resizebox{1\linewidth}{!}{
    \begin{tabular}{c|c|cccc|cccc|cccc}
    \hline
          \multicolumn{2}{c|}{Datasets} &  \multicolumn{4}{c}{MOOC}& \multicolumn{4}{c}{Amazon} & \multicolumn{4}{c}{Gowalla}\\\hline
          Metrics& Backbones& R@20& R@50& N@20& N@50 & R@20& R@50& N@20& N@50& R@20& R@50& N@20& N@50 \\\hline
          \multirow{2}{*}{SimGCL} & LightGCN & 0.3503 & 0.5032 & 0.2109 & 0.2428 & 0.0462 & 0.0837 & 0.0230 & 0.0327 & 0.2028& 0.3126& 0.1184& 0.1459 \\
          ~ & FourierKAN-GCF & \textbf{0.3639} & \textbf{0.5188} & \textbf{0.2216} & \textbf{0.2543} & \textbf{0.0482} & \textbf{0.0868} & \textbf{0.0258} & \textbf{0.0352} & \textbf{0.2084}& \textbf{0.3163}& \textbf{0.1234}& \textbf{0.1521} \\\hline
          \multirow{2}{*}{LightGCL} & LightGCN & 0.2742 & 0.4139 & 0.1697 & 0.2018 & 0.0453 & 0.0818 & 0.0225 & 0.0318 & 0.1888& 0.2976& 0.1081& 0.1348 \\
          ~ & FourierKAN-GCF & \textbf{0.2823} & \textbf{0.4291} & \textbf{0.1733} & \textbf{0.2072} & \textbf{0.0479} & \textbf{0.0864} & \textbf{0.0250} & \textbf{0.0347} & \textbf{0.2007}& \textbf{0.3100}& \textbf{0.1198}& \textbf{0.1452} \\\hline
          \multirow{2}{*}{RecDCL} & LightGCN & 0.3531 & 0.5009 & 0.2113 & 0.2423 & 0.0469 & 0.0848 & 0.0235 & 0.0334 & 0.1993& 0.3052& 0.1160& 0.1433 \\
          ~ & FourierKAN-GCF & \textbf{0.3608} & \textbf{0.5120} & \textbf{0.2175} & \textbf{0.2514} & \textbf{0.0484} & \textbf{0.0863} & \textbf{0.0253} & \textbf{0.0349} & \textbf{0.2051}& \textbf{0.3113}& \textbf{0.1203}& \textbf{0.1472} \\\hline
          % \hline
    \end{tabular}
    }
    % \\ $^1$Datasets can be accessed at \href{http://moocdata.cn}{http://moocdata.cn}, \href{http://jmcauley.ucsd.edu/data/amazon/links.html}{http://jmcauley.ucsd.edu/data/amazon/links.html}, and \href{https://snap.stanford.edu/data/loc-gowalla.html}{https://snap.stanford.edu/data/loc-gowalla.html}.
\vskip -0.15in
\end{table*}

We conduct experiments on three real-world datasets: MOOC, Amazon Video Games (Amazon), and Gowalla. Details can be found in Table~\ref{tab:dataset_statistics}. For a fair comparison, we sort all observed user-item interactions chronologically based on the interaction timestamps. Then, we split each dataset with a ratio of 7:1:2 for training, validation, and testing. Regarding evaluation metrics, we adopt two well-established metrics \cite{jarvelin2002cumulated}: Recall@K (R@K) and NDCG@K (N@K). 
% We report the average metrics of all users in the test dataset under $K$ = 20 and $K$ = 50.

\subsection{Baselines and Experimental Settings}
To verify the effectiveness of FourierKAN-GCF, we select five GCN-based backbone models (NGCF \cite{wang2019neural}, LR-GCCF \cite{chen2020revisiting}, LightGCN \cite{he2020lightgcn}, UltraGCN \cite{mao2021ultragcn}, and IMP-GCN \cite{liu2021interest}) as baselines. Moreover, FourierKAN-GCF can also be a more powerful backbone to enhance existing advanced self-supervised enhanced graph-based models (SimGCL \cite{yu2022graph}, LightGCL \cite{cailightgcl}, and RecDCL \cite{zhang2024recdcl}). For a fair comparison, we fix the embedding size of both users and items to 64 for all models, initialize embedding parameters with the Xavier initialization \cite{glorot2010understanding}, and use Adam \cite{kingma2014adam} as optimizer. Besides, we tune the hyper-parameters of each baseline following their published papers. For FourierKAN-GCF, we fix the $\lambda$ = $1$ for $L_2$ regularization, and tune layer number $L$ from 1 to 4. The gridsize $g$ is searched from \{1, 2, 4, 8\}. Message and node dropout ratios $p_m$ and $p_n$ is searched from \{0.0, 0.1, 0.2, 0.3\}.

% To verify the effectiveness of FourierKAN-GCF, we select two matrix factorization models (BPR-MF \cite{rendle2012bpr} and BUIR \cite{lee2021bootstrapping}) and six advanced GCN-based models (NGCF \cite{wang2019neural}, LR-GCCF \cite{chen2020revisiting}, LightGCN \cite{he2020lightgcn}, UltraGCN \cite{mao2021ultragcn}, IMP-GCN \cite{liu2021interest}, and LightGCL \cite{cailightgcl}) for comparison. For a fair comparison, we fix the embedding size of both users and items to 64 for all models, initialize the embedding parameters with the Xavier initialization \cite{glorot2010understanding}, and use Adam \cite{kingma2014adam} as the optimizer. Besides, we carefully tune the hyper-parameters of each baseline following their published papers to find optimal settings. For FourierKAN-GCF, we tune the $\lambda$ of $L_2$ regularization term in \{0, 1e-2, 1e-1, 1, 10\}, and layer number $L$ from 1 to 4. The gridsize $g$ is searched from \{1, 2, 4, 8\}. We further grid search the message and node dropout ratios $p_m$ and $p_n$ in \{0.0, 0.1, 0.2, 0.3\}.

\subsection{Performance Comparison} 
The results of our experiments are listed in Table~\ref{tab:results}. Our FourierKAN-GCF outperforms all baselines in three datasets across various metrics. Moreover, FourierKAN-GCF is second only to LightGCN in efficiency, which demonstrates it effectively reduces the training difficulty associated with feature transformation. We owe our superiority to FourierKAN, which is easier to train and has greater representation power than MLP. Note that FourierKAN-GCF can adjust the training difficulty by adjusting the grid size $g$.

% The results of our experiments are listed in Table~\ref{tab:results}. We can identify that the majority of GCN-based methods outperform traditional matrix factorization methods, which is a common trend. Our FourierKAN-GCF outperforms all baselines in three datasets across various metrics. We owe our superiority to FourierKAN, which is easier to train and has greater representation power than MLP. Note that FourierKAN-GCF can adjust the training difficulty by adjusting the grid size $g$.

\subsection{Compatibility Analysis}
We further evaluate the compatibility of FourierKAN-GCF with advanced self-supervised graph-based models by replacing their backbones. As shown in Table~\ref{tab:results2}, adopting FourierKAN-GCF significantly improves performance, highlighting the importance of feature transformation in GCF and establishing FourierKAN-GCF as an effective solution.

\subsection{Ablation Study}
Table~\ref{tab:results} also demonstrates the significance of dropout strategies. We use w/o MD and w/o ND to denote without message dropout and without node dropout, respectively. This ablation study shows that both message dropout and node dropout play a distinct role in improving model representation power and model robustness. We use KAN-GCF to denote a variant that utilizes our standard KAN function to replace transform matrix $\mathbf{W}_2$ in NGCF. It shows that standard KAN is better than MLP but worse than our Fourier KAN.

\begin{figure}[h]
\vskip -0.1in
    \centering
    \subfigure[Layer number $L$ (Recall@20)] {
        \label{fig:l}
        \includegraphics[width=0.48\linewidth]{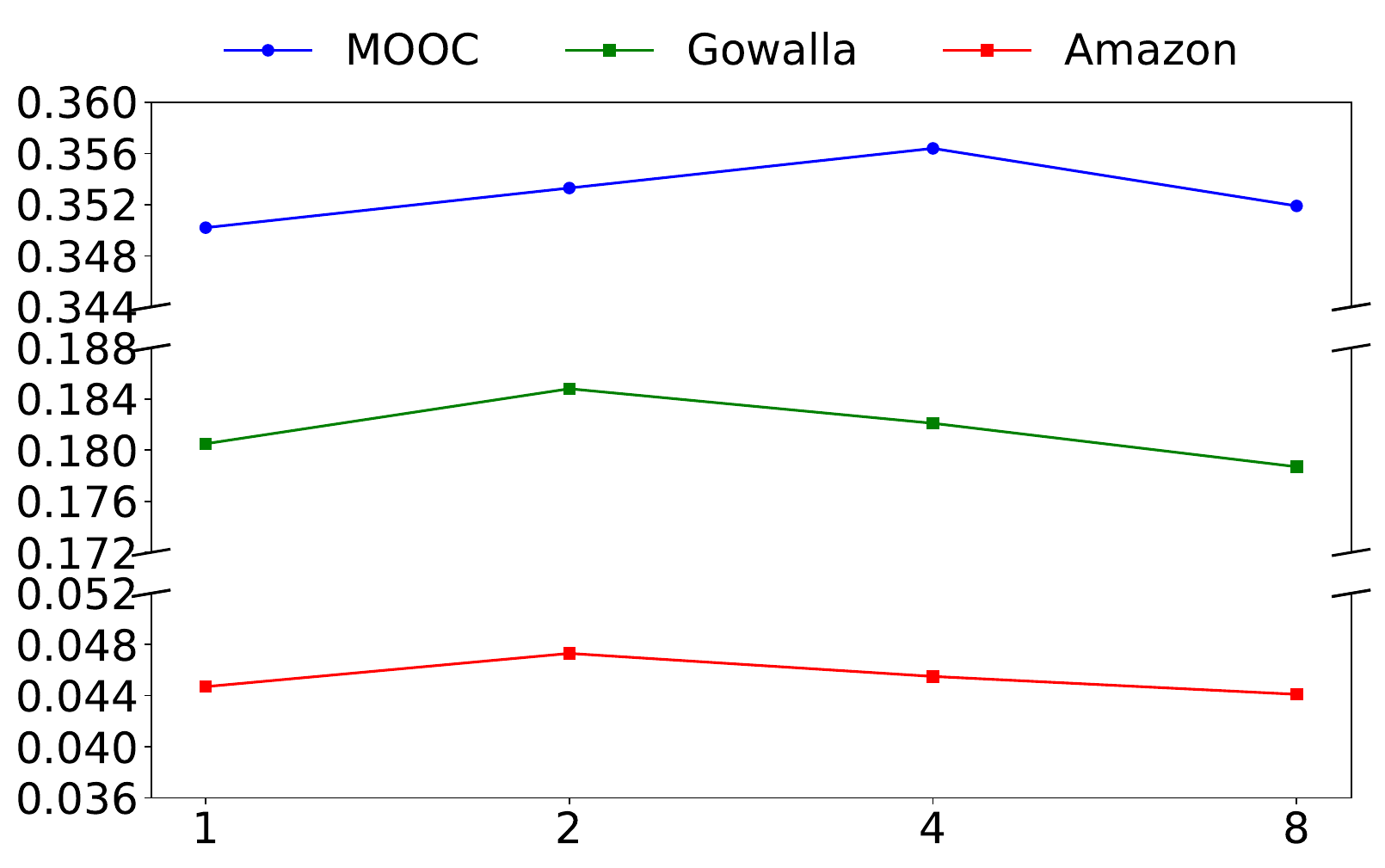}
        }   \hspace{-0.1in}
    \subfigure[Grid size $g$ (Recall@20)] {
        \label{fig:g}
        \includegraphics[width=0.48\linewidth]{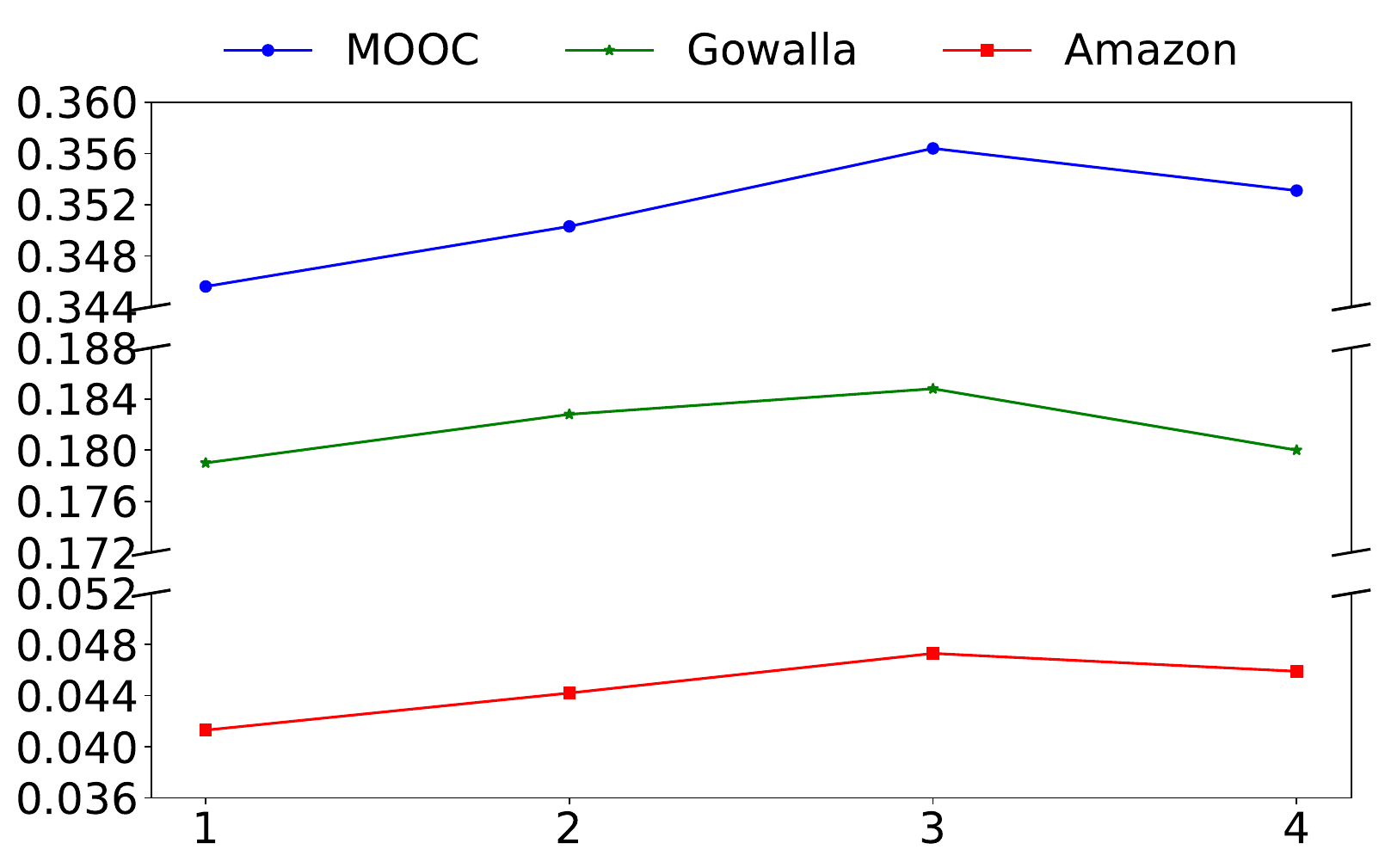}
        }  
    \vskip -0.2in
    \caption{Effective of layer number $L$ and grid size $g$.}   
    \label{fig:efficiency}
    \vskip -0.2in
\end{figure}

\begin{figure}[h]
\vskip -0.05in
    \centering
    \subfigure[MOOC] {
        \label{fig:mooc}
        \includegraphics[width=0.32\linewidth]{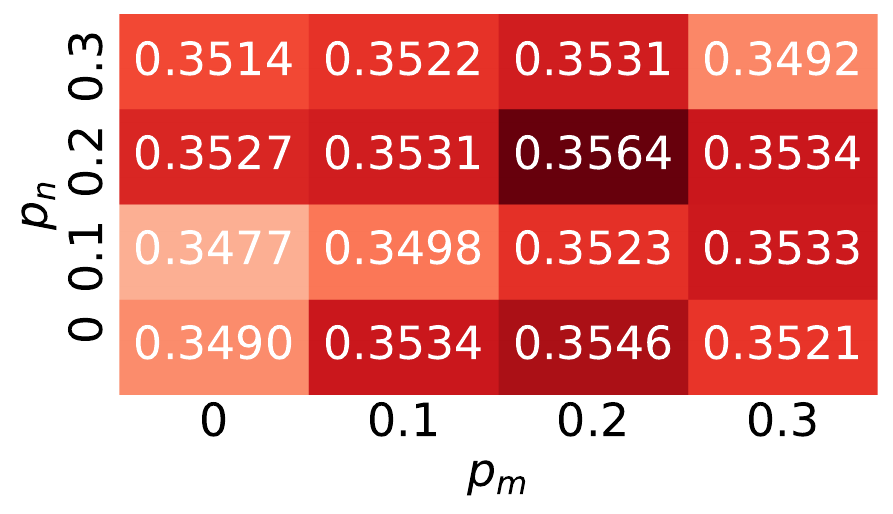}
        }  \hspace{-0.1in}
    \subfigure[Amazon] {
        \label{fig:amazon}
        \includegraphics[width=0.32\linewidth]{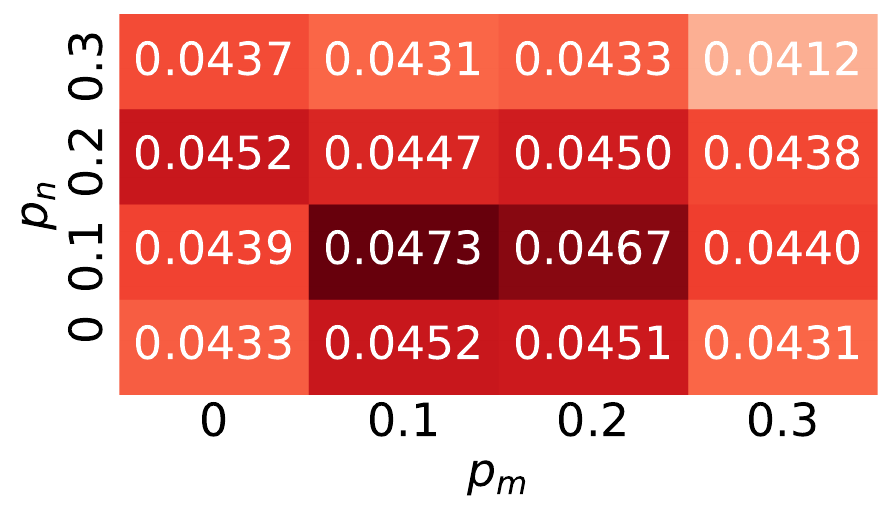}
        }  \hspace{-0.1in}
    \subfigure[Gowalla] {
        \label{fig:gowalla}
        \includegraphics[width=0.32\linewidth]{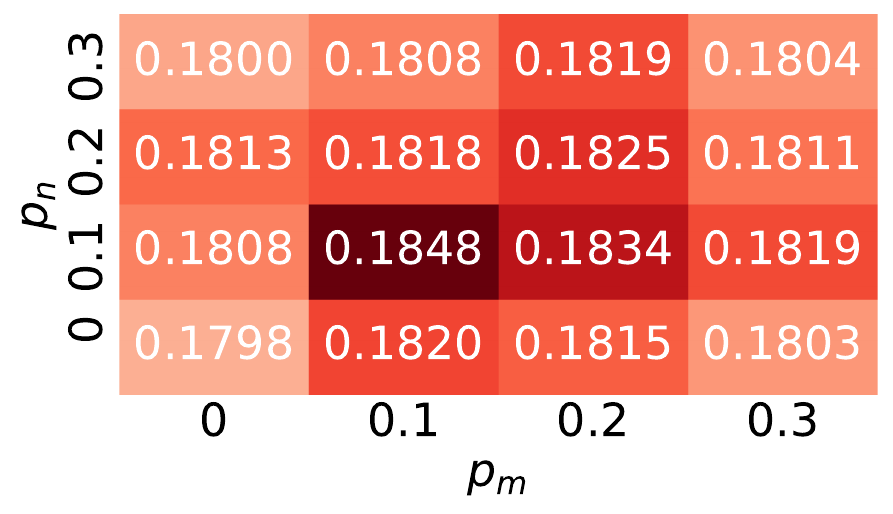}
        }
    \vskip -0.2in
    \caption{Study of dropout ratio pair $p_m$ and $p_n$ (Recall@20).}   
    \label{fig:dropout}
    \vskip -0.25in
\end{figure}

\subsection{Sensitivity Analysis}
To analyze the hyper-parameter sensitivity of FourierKAN-GCF, we test the performance of FourierKAN-GCF on three datasets with different hyper-parameters. The optimal layer number $L$ is 3 on all datasets, shown in Figure~\ref{fig:l}. Besides, Figure~\ref{fig:g} demonstrates that for MOOC dataset, the optimal ($p_m$, $p_n$) pair is (0.2, 0.2), and for Amazon and Gowalla datasets, the optimal pairs are all (0.1, 0.1). As Figure~\ref{fig:dropout} illustrated, for the relatively dense dataset MOOC, $g$ = 4 is the best grid size. For the relatively sparse datasets Amazon and Gowalla, $g$ = 2 is the best grid size. This further demonstrates that FourierKAN can be easily adapted to various datasets.
% \subsubsection{Effect of the Layer Number $L$}
% \subsubsection{Effect of the Dropout Ratio $p_m$ and $p_n$}
% \subsubsection{Effect of the Gridsize $g$ in Fourier KAN}
\section{Conclusion}
In this paper, we revisit feature transformation and nonlinear operations in the message-passing mechanism of GCNs. While feature transformation enhances interaction representation and boosts performance, it increases training complexity. To this end, we introduce a new feature transformation for GCF called FourierKAN. Inspired by KAN, FourierKAN employs the Fourier Coefficients instead of the Spline function in standard KAN. We further propose a simple yet effective GCF model (FourierKAN-GCF), which reduces the difficulty of training. In addition, FourierKAN-GCF can be integrated into existing advanced self-supervised models as a backbone, replacing their original backbone to achieve enhanced performance. Extensive experiments on public datasets verify the superiority of our model over the advanced methods.

\section*{GenAI Usage Disclosure}
No GenAI tools were used in any stage of the research, nor in the writing.

% \newpage

\begin{acks}
This work was supported by the Hong Kong UGC General Research Fund no. 17203320 and 17209822, and the project grants from the HKU-SCF FinTech Academy.
\end{acks}

\bibliographystyle{ACM-Reference-Format}
\balance
% \bibliography{reference}

\begin{thebibliography}{28}

%%% the Web address.

\ifx \showCODEN    \undefined \def \showCODEN     #1{\unskip}     \fi
\ifx \showDOI      \undefined \def \showDOI       #1{#1}\fi
\ifx \showISBNx    \undefined \def \showISBNx     #1{\unskip}     \fi
\ifx \showISBNxiii \undefined \def \showISBNxiii  #1{\unskip}     \fi
\ifx \showISSN     \undefined \def \showISSN      #1{\unskip}     \fi
\ifx \showLCCN     \undefined \def \showLCCN      #1{\unskip}     \fi
\ifx \shownote     \undefined \def \shownote      #1{#1}          \fi
\ifx \showarticletitle \undefined \def \showarticletitle #1{#1}   \fi
\ifx \showURL      \undefined \def \showURL       {\relax}        \fi
% invisible to TeX
\providecommand\bibfield[2]{#2}
\providecommand\bibinfo[2]{#2}
\providecommand\natexlab[1]{#1}
\providecommand\showeprint[2][]{arXiv:#2}

\bibitem[Cai et~al\mbox{.}(2023)]%
        {cailightgcl}
\bibfield{author}{\bibinfo{person}{Xuheng Cai}, \bibinfo{person}{Chao Huang}, \bibinfo{person}{Lianghao Xia}, {and} \bibinfo{person}{Xubin Ren}.} \bibinfo{year}{2023}\natexlab{}.
\newblock \showarticletitle{LightGCL: Simple Yet Effective Graph Contrastive Learning for Recommendation}. In \bibinfo{booktitle}{\emph{The Eleventh International Conference on Learning Representations}}.
\newblock


\bibitem[Chen et~al\mbox{.}(2020)]%
        {chen2020revisiting}
\bibfield{author}{\bibinfo{person}{Lei Chen}, \bibinfo{person}{Le Wu}, \bibinfo{person}{Richang Hong}, \bibinfo{person}{Kun Zhang}, {and} \bibinfo{person}{Meng Wang}.} \bibinfo{year}{2020}\natexlab{}.
\newblock \showarticletitle{Revisiting graph based collaborative filtering: A linear residual graph convolutional network approach}. In \bibinfo{booktitle}{\emph{Proceedings of the AAAI conference on artificial intelligence}}, Vol.~\bibinfo{volume}{34}. \bibinfo{pages}{27--34}.
\newblock


\bibitem[Chen et~al\mbox{.}(2025a)]%
        {chen2025don}
\bibfield{author}{\bibinfo{person}{Zheyu Chen}, \bibinfo{person}{Jinfeng Xu}, {and} \bibinfo{person}{Haibo Hu}.} \bibinfo{year}{2025}\natexlab{a}.
\newblock \showarticletitle{Don’t Lose Yourself: Boosting Multimodal Recommendation via Reducing Node-neighbor Discrepancy in Graph Convolutional Network}. In \bibinfo{booktitle}{\emph{ICASSP 2025-2025 IEEE International Conference on Acoustics, Speech and Signal Processing (ICASSP)}}. IEEE, \bibinfo{pages}{1--5}.
\newblock


\bibitem[Chen et~al\mbox{.}(2025b)]%
        {chen2025squeeze}
\bibfield{author}{\bibinfo{person}{Zheyu Chen}, \bibinfo{person}{Jinfeng Xu}, \bibinfo{person}{Yutong Wei}, {and} \bibinfo{person}{Ziyue Peng}.} \bibinfo{year}{2025}\natexlab{b}.
\newblock \showarticletitle{Squeeze and Excitation: A Weighted Graph Contrastive Learning for Collaborative Filtering}. In \bibinfo{booktitle}{\emph{Proceedings of the 48th International ACM SIGIR Conference on Research and Development in Information Retrieval}}. \bibinfo{pages}{2769--2773}.
\newblock


\bibitem[Cybenko(1989)]%
        {cybenko1989approximation}
\bibfield{author}{\bibinfo{person}{George Cybenko}.} \bibinfo{year}{1989}\natexlab{}.
\newblock \showarticletitle{Approximation by superpositions of a sigmoidal function}.
\newblock \bibinfo{journal}{\emph{Mathematics of control, signals and systems}} \bibinfo{volume}{2}, \bibinfo{number}{4} (\bibinfo{year}{1989}), \bibinfo{pages}{303--314}.
\newblock


\bibitem[Glorot and Bengio(2010)]%
        {glorot2010understanding}
\bibfield{author}{\bibinfo{person}{Xavier Glorot} {and} \bibinfo{person}{Yoshua Bengio}.} \bibinfo{year}{2010}\natexlab{}.
\newblock \showarticletitle{Understanding the difficulty of training deep feedforward neural networks}. In \bibinfo{booktitle}{\emph{Proceedings of the thirteenth international conference on artificial intelligence and statistics}}. JMLR Workshop and Conference Proceedings, \bibinfo{pages}{249--256}.
\newblock


\bibitem[He et~al\mbox{.}(2020)]%
        {he2020lightgcn}
\bibfield{author}{\bibinfo{person}{Xiangnan He}, \bibinfo{person}{Kuan Deng}, \bibinfo{person}{Xiang Wang}, \bibinfo{person}{Yan Li}, \bibinfo{person}{Yongdong Zhang}, {and} \bibinfo{person}{Meng Wang}.} \bibinfo{year}{2020}\natexlab{}.
\newblock \showarticletitle{Lightgcn: Simplifying and powering graph convolution network for recommendation}. In \bibinfo{booktitle}{\emph{Proceedings of the 43rd International ACM SIGIR conference on research and development in Information Retrieval}}. \bibinfo{pages}{639--648}.
\newblock


\bibitem[He et~al\mbox{.}(2017)]%
        {he2017neural}
\bibfield{author}{\bibinfo{person}{Xiangnan He}, \bibinfo{person}{Lizi Liao}, \bibinfo{person}{Hanwang Zhang}, \bibinfo{person}{Liqiang Nie}, \bibinfo{person}{Xia Hu}, {and} \bibinfo{person}{Tat-Seng Chua}.} \bibinfo{year}{2017}\natexlab{}.
\newblock \showarticletitle{Neural collaborative filtering}. In \bibinfo{booktitle}{\emph{Proceedings of the 26th international conference on world wide web}}. \bibinfo{pages}{173--182}.
\newblock


\bibitem[J{\"a}rvelin and Kek{\"a}l{\"a}inen(2002)]%
        {jarvelin2002cumulated}
\bibfield{author}{\bibinfo{person}{Kalervo J{\"a}rvelin} {and} \bibinfo{person}{Jaana Kek{\"a}l{\"a}inen}.} \bibinfo{year}{2002}\natexlab{}.
\newblock \showarticletitle{Cumulated gain-based evaluation of IR techniques}.
\newblock \bibinfo{journal}{\emph{ACM Transactions on Information Systems (TOIS)}} \bibinfo{volume}{20}, \bibinfo{number}{4} (\bibinfo{year}{2002}), \bibinfo{pages}{422--446}.
\newblock


\bibitem[Jiang et~al\mbox{.}(2023)]%
        {jiang2023adaptive}
\bibfield{author}{\bibinfo{person}{Yangqin Jiang}, \bibinfo{person}{Chao Huang}, {and} \bibinfo{person}{Lianghao Huang}.} \bibinfo{year}{2023}\natexlab{}.
\newblock \showarticletitle{Adaptive graph contrastive learning for recommendation}. In \bibinfo{booktitle}{\emph{Proceedings of the 29th ACM SIGKDD conference on knowledge discovery and data mining}}. \bibinfo{pages}{4252--4261}.
\newblock


\bibitem[Kingma and Ba(2015)]%
        {kingma2014adam}
\bibfield{author}{\bibinfo{person}{Diederik~P. Kingma} {and} \bibinfo{person}{Jimmy Ba}.} \bibinfo{year}{2015}\natexlab{}.
\newblock \showarticletitle{Adam: {A} Method for Stochastic Optimization}. In \bibinfo{booktitle}{\emph{Proceedings of the 3rd International Conference on Learning Representations}}.
\newblock


\bibitem[Kolmogorov(1961)]%
        {kolmogorov1961representation}
\bibfield{author}{\bibinfo{person}{Andrey~Nikolaevich Kolmogorov}.} \bibinfo{year}{1961}\natexlab{}.
\newblock \bibinfo{booktitle}{\emph{On the representation of continuous functions of several variables by superpositions of continuous functions of a smaller number of variables}}.
\newblock \bibinfo{publisher}{American Mathematical Society}.
\newblock


\bibitem[Liu et~al\mbox{.}(2021)]%
        {liu2021interest}
\bibfield{author}{\bibinfo{person}{Fan Liu}, \bibinfo{person}{Zhiyong Cheng}, \bibinfo{person}{Lei Zhu}, \bibinfo{person}{Zan Gao}, {and} \bibinfo{person}{Liqiang Nie}.} \bibinfo{year}{2021}\natexlab{}.
\newblock \showarticletitle{Interest-aware message-passing GCN for recommendation}. In \bibinfo{booktitle}{\emph{Proceedings of the web conference 2021}}. \bibinfo{pages}{1296--1305}.
\newblock


\bibitem[Liu et~al\mbox{.}(2024)]%
        {liu2024kan}
\bibfield{author}{\bibinfo{person}{Ziming Liu}, \bibinfo{person}{Yixuan Wang}, \bibinfo{person}{Sachin Vaidya}, \bibinfo{person}{Fabian Ruehle}, \bibinfo{person}{James Halverson}, \bibinfo{person}{Marin Solja{\v{c}}i{\'c}}, \bibinfo{person}{Thomas~Y Hou}, {and} \bibinfo{person}{Max Tegmark}.} \bibinfo{year}{2024}\natexlab{}.
\newblock \showarticletitle{Kan: Kolmogorov-arnold networks}.
\newblock \bibinfo{journal}{\emph{arXiv preprint arXiv:2404.19756}} (\bibinfo{year}{2024}).
\newblock


\bibitem[Mao et~al\mbox{.}(2021)]%
        {mao2021ultragcn}
\bibfield{author}{\bibinfo{person}{Kelong Mao}, \bibinfo{person}{Jieming Zhu}, \bibinfo{person}{Xi Xiao}, \bibinfo{person}{Biao Lu}, \bibinfo{person}{Zhaowei Wang}, {and} \bibinfo{person}{Xiuqiang He}.} \bibinfo{year}{2021}\natexlab{}.
\newblock \showarticletitle{UltraGCN: ultra simplification of graph convolutional networks for recommendation}. In \bibinfo{booktitle}{\emph{Proceedings of the 30th ACM international conference on information \& knowledge management}}. \bibinfo{pages}{1253--1262}.
\newblock


\bibitem[Nussbaumer and Nussbaumer(1982)]%
        {nussbaumer1982fast}
\bibfield{author}{\bibinfo{person}{Henri~J Nussbaumer} {and} \bibinfo{person}{Henri~J Nussbaumer}.} \bibinfo{year}{1982}\natexlab{}.
\newblock \bibinfo{booktitle}{\emph{The fast Fourier transform}}.
\newblock \bibinfo{publisher}{Springer}.
\newblock


\bibitem[Rendle et~al\mbox{.}(2009)]%
        {rendle2009bpr}
\bibfield{author}{\bibinfo{person}{Steffen Rendle}, \bibinfo{person}{Christoph Freudenthaler}, \bibinfo{person}{Zeno Gantner}, {and} \bibinfo{person}{Lars Schmidt-Thieme}.} \bibinfo{year}{2009}\natexlab{}.
\newblock \showarticletitle{BPR: Bayesian personalized ranking from implicit feedback}. In \bibinfo{booktitle}{\emph{Proceedings of the Twenty-Fifth Conference on Uncertainty in Artificial Intelligence}}. \bibinfo{pages}{452--461}.
\newblock


\bibitem[Wang et~al\mbox{.}(2019)]%
        {wang2019neural}
\bibfield{author}{\bibinfo{person}{Xiang Wang}, \bibinfo{person}{Xiangnan He}, \bibinfo{person}{Meng Wang}, \bibinfo{person}{Fuli Feng}, {and} \bibinfo{person}{Tat-Seng Chua}.} \bibinfo{year}{2019}\natexlab{}.
\newblock \showarticletitle{Neural graph collaborative filtering}. In \bibinfo{booktitle}{\emph{Proceedings of the 42nd international ACM SIGIR conference on Research and development in Information Retrieval}}. \bibinfo{pages}{165--174}.
\newblock


\bibitem[Wang et~al\mbox{.}(2024)]%
        {wang2024expressiveness}
\bibfield{author}{\bibinfo{person}{Yixuan Wang}, \bibinfo{person}{Jonathan~W Siegel}, \bibinfo{person}{Ziming Liu}, {and} \bibinfo{person}{Thomas~Y Hou}.} \bibinfo{year}{2024}\natexlab{}.
\newblock \showarticletitle{On the expressiveness and spectral bias of KANs}.
\newblock \bibinfo{journal}{\emph{arXiv preprint arXiv:2410.01803}} (\bibinfo{year}{2024}).
\newblock


\bibitem[Xu et~al\mbox{.}(2025a)]%
        {xu2025mdvt}
\bibfield{author}{\bibinfo{person}{Jinfeng Xu}, \bibinfo{person}{Zheyu Chen}, \bibinfo{person}{Jinze Li}, \bibinfo{person}{Shuo Yang}, \bibinfo{person}{Hewei Wang}, \bibinfo{person}{Yijie Li}, \bibinfo{person}{Mengran Li}, \bibinfo{person}{Puzhen Wu}, {and} \bibinfo{person}{Edith~CH Ngai}.} \bibinfo{year}{2025}\natexlab{a}.
\newblock \showarticletitle{MDVT: Enhancing Multimodal Recommendation with Model-Agnostic Multimodal-Driven Virtual Triplets}.
\newblock \bibinfo{journal}{\emph{arXiv preprint arXiv:2505.16665}} (\bibinfo{year}{2025}).
\newblock


\bibitem[Xu et~al\mbox{.}(2024a)]%
        {xu2024aligngroup}
\bibfield{author}{\bibinfo{person}{Jinfeng Xu}, \bibinfo{person}{Zheyu Chen}, \bibinfo{person}{Jinze Li}, \bibinfo{person}{Shuo Yang}, \bibinfo{person}{Hewei Wang}, {and} \bibinfo{person}{Edith~CH Ngai}.} \bibinfo{year}{2024}\natexlab{a}.
\newblock \showarticletitle{AlignGroup: Learning and Aligning Group Consensus with Member Preferences for Group Recommendation}. In \bibinfo{booktitle}{\emph{Proceedings of the 33rd ACM International Conference on Information and Knowledge Management}}. \bibinfo{pages}{2682--2691}.
\newblock


\bibitem[Xu et~al\mbox{.}(2024b)]%
        {xu2024improving}
\bibfield{author}{\bibinfo{person}{Jinfeng Xu}, \bibinfo{person}{Zheyu Chen}, \bibinfo{person}{Zixiao Ma}, \bibinfo{person}{Jiyi Liu}, {and} \bibinfo{person}{Edith~CH Ngai}.} \bibinfo{year}{2024}\natexlab{b}.
\newblock \showarticletitle{Improving Consumer Experience With Pre-Purify Temporal-Decay Memory-Based Collaborative Filtering Recommendation for Graduate School Application}.
\newblock \bibinfo{journal}{\emph{IEEE Transactions on Consumer Electronics}} (\bibinfo{year}{2024}).
\newblock


\bibitem[Xu et~al\mbox{.}(2025b)]%
        {xu2025cohesion}
\bibfield{author}{\bibinfo{person}{Jinfeng Xu}, \bibinfo{person}{Zheyu Chen}, \bibinfo{person}{Wei Wang}, \bibinfo{person}{Xiping Hu}, \bibinfo{person}{Sang-Wook Kim}, {and} \bibinfo{person}{Edith~CH Ngai}.} \bibinfo{year}{2025}\natexlab{b}.
\newblock \showarticletitle{COHESION: Composite Graph Convolutional Network with Dual-Stage Fusion for Multimodal Recommendation}. In \bibinfo{booktitle}{\emph{Proceedings of the 48th International ACM SIGIR Conference on Research and Development in Information Retrieval}}. \bibinfo{pages}{1830--1839}.
\newblock


\bibitem[Xu et~al\mbox{.}(2025c)]%
        {xu2025best}
\bibfield{author}{\bibinfo{person}{Jinfeng Xu}, \bibinfo{person}{Zheyu Chen}, \bibinfo{person}{Shuo Yang}, \bibinfo{person}{Jinze Li}, {and} \bibinfo{person}{Edith~CH Ngai}.} \bibinfo{year}{2025}\natexlab{c}.
\newblock \showarticletitle{The Best is Yet to Come: Graph Convolution in the Testing Phase for Multimodal Recommendation}.
\newblock \bibinfo{journal}{\emph{arXiv preprint arXiv:2507.18489}} (\bibinfo{year}{2025}).
\newblock


\bibitem[Xu et~al\mbox{.}(2025d)]%
        {xu2025mentor}
\bibfield{author}{\bibinfo{person}{Jinfeng Xu}, \bibinfo{person}{Zheyu Chen}, \bibinfo{person}{Shuo Yang}, \bibinfo{person}{Jinze Li}, \bibinfo{person}{Hewei Wang}, {and} \bibinfo{person}{Edith~CH Ngai}.} \bibinfo{year}{2025}\natexlab{d}.
\newblock \showarticletitle{Mentor: multi-level self-supervised learning for multimodal recommendation}. In \bibinfo{booktitle}{\emph{Proceedings of the AAAI Conference on Artificial Intelligence}}, Vol.~\bibinfo{volume}{39}. \bibinfo{pages}{12908--12917}.
\newblock


\bibitem[Xu et~al\mbox{.}(2025e)]%
        {xu2025survey}
\bibfield{author}{\bibinfo{person}{Jinfeng Xu}, \bibinfo{person}{Zheyu Chen}, \bibinfo{person}{Shuo Yang}, \bibinfo{person}{Jinze Li}, \bibinfo{person}{Wei Wang}, \bibinfo{person}{Xiping Hu}, \bibinfo{person}{Steven Hoi}, {and} \bibinfo{person}{Edith Ngai}.} \bibinfo{year}{2025}\natexlab{e}.
\newblock \showarticletitle{A Survey on Multimodal Recommender Systems: Recent Advances and Future Directions}.
\newblock \bibinfo{journal}{\emph{arXiv preprint arXiv:2502.15711}} (\bibinfo{year}{2025}).
\newblock


\bibitem[Yu et~al\mbox{.}(2022)]%
        {yu2022graph}
\bibfield{author}{\bibinfo{person}{Junliang Yu}, \bibinfo{person}{Hongzhi Yin}, \bibinfo{person}{Xin Xia}, \bibinfo{person}{Tong Chen}, \bibinfo{person}{Lizhen Cui}, {and} \bibinfo{person}{Quoc Viet~Hung Nguyen}.} \bibinfo{year}{2022}\natexlab{}.
\newblock \showarticletitle{Are graph augmentations necessary? simple graph contrastive learning for recommendation}. In \bibinfo{booktitle}{\emph{Proceedings of the 45th international ACM SIGIR conference on research and development in information retrieval}}. \bibinfo{pages}{1294--1303}.
\newblock


\bibitem[Zhang et~al\mbox{.}(2024)]%
        {zhang2024recdcl}
\bibfield{author}{\bibinfo{person}{Dan Zhang}, \bibinfo{person}{Yangliao Geng}, \bibinfo{person}{Wenwen Gong}, \bibinfo{person}{Zhongang Qi}, \bibinfo{person}{Zhiyu Chen}, \bibinfo{person}{Xing Tang}, \bibinfo{person}{Ying Shan}, \bibinfo{person}{Yuxiao Dong}, {and} \bibinfo{person}{Jie Tang}.} \bibinfo{year}{2024}\natexlab{}.
\newblock \showarticletitle{RecDCL: Dual Contrastive Learning for Recommendation}. In \bibinfo{booktitle}{\emph{Proceedings of the ACM on Web Conference 2024}}. \bibinfo{pages}{3655--3666}.
\newblock


\end{thebibliography}

\appendix

\end{document}